\documentclass[runningheads]{llncs}
\usepackage[T1]{fontenc}
\usepackage{graphicx}
\usepackage{booktabs}
\usepackage{multirow}
\usepackage{enumitem}
\usepackage{amsmath}
\usepackage{amssymb}
\usepackage{xspace}

\PassOptionsToPackage{hyphens}{url}
\usepackage{hyperref}
\usepackage{color}

\newcommand{\ModelName}{GNRR\xspace}
\begin{document}
\title{Cross-Document Neural Re-Ranking via Query-Induced Subgraphs\thanks{%
Author's version. Accepted at AIxIA 2026. The version of record will be published by Springer in LNAI.}}
%
%
\author{
Andrea Giuseppe Di Francesco\inst{1,2,}\thanks{These authors equally contributed to the paper}
\and Christian Giannetti\inst{1,3,*}
\and Federico Siciliano\inst{4}
\and Nicola Tonellotto\inst{5}
\and Fabrizio Silvestri\inst{1}
}

\authorrunning{Di Francesco et al.}

\institute{
Sapienza University of Rome, Rome, Italy
\email{\{difrancesco,giannetti,fsilvestri\}@diag.uniroma1.it}
\and
ISTI-CNR, Pisa, Italy
\and
University of Cambridge, Cambridge, UK
\and
Istituto Nazionale di Geofisica e Vulcanologia, Rome, Italy
\email{federico.siciliano@ingv.it}
\and
University of Pisa, Pisa, Italy
\email{nicola.tonellotto@unipi.it}\\
}
\maketitle
\begin{abstract}
Neural re-rankers typically score query-document pairs independently, neglecting cross-document context within the retrieved candidate set.
We propose Graph Neural Re-Ranking (\ModelName), a framework that extracts a sparse, query-induced subgraph from a pre-computed semantic corpus graph and applies Graph Neural Networks (GNN) to propagate cross-document signals.
Unlike self-attention re-rankers, which scale quadratically with the number of candidates ($\mathcal{O}(K^2)$), \ModelName achieves $\mathcal{O}(c \cdot K)$ online complexity, where $c$ is the fixed corpus graph degree and $K$ the candidate set size.
We evaluate five GNN operators within this framework and find that architecture choice substantially affects generalization to harder queries: the GCN variant is the only one that consistently improves over TCT-ColBERT across all three TREC benchmarks.
On TREC-DLHard, the most challenging evaluation benchmark, \ModelName achieves $+5.2\%$ relative AP over TCT-ColBERT and $+9.0\%$ AP over a self-attention re-ranker.
Notably, self-attention re-ranking \emph{degrades} AP on DLHard ($-3.5\%$ versus TCT-ColBERT), suggesting that sparse corpus-graph structure provides a complementary re-ranking signal that dense self-attention fails to capture.
Efficiency analysis shows that GNN models require fewer parameters and lower per-query latency at $K=1000$ than self-attention, with linear rather than quadratic scaling in candidate set size.
Code to reproduce our experiment is available at \url{https://github.com/difra100/Graph-Neural-Re-Ranking-via-Corpus-Graph}
\end{abstract}


\section{Introduction}
Modern information retrieval systems typically follow a two-stage pipeline: an initial retrieval step (e.g., BM25~\cite{bm25} or dense retrieval~\cite{colbert}) generates a set of candidate documents, which is refined by a neural re-ranking model (e.g., a transformer-based scorer~\cite{tctcolbert,vaswani2017attention}) to produce the final ranking.
Despite their effectiveness, most neural re-rankers score each query-document pair independently, discarding relationships among candidates in the retrieved set~\cite{monobert,RepBert}.

Capturing cross-document context is desirable but costly. Prior work~\cite{pang2020setrank,pobrotyn2020context,pasumarthi2020permutation} employs self-attention mechanisms over the candidate set, treating documents as a permutation-invariant set~\cite{lee2019set,pang2020setrank}. However, self-attention incurs quadratic cost $\mathcal{O}(K^2)$ in the number of documents, limiting its applicability in latency-constrained settings.
Moreover, as our experiments show, self-attention re-ranking can \emph{degrade} performance on difficult queries where relevant documents rank low among predominantly non-relevant candidates.

We take a different approach. Rather than attending to all candidate pairs, \ModelName builds a sparse, query-induced subgraph from the retrieved top-$K$ documents using a pre-computed semantic corpus graph~\cite{corpusgraph} and applies lightweight Graph Neural Networks to propagate cross-document signals. GNN message passing costs $\mathcal{O}(c \cdot K)$ online, where $c$ is the fixed corpus graph degree, linear in $K$ compared to $\mathcal{O}(K^2)$ for self-attention. Because the corpus graph is pre-computed offline, the online overhead is substantially reduced compared to dense attention-based approaches.

Our contributions are as follows:
\begin{itemize}
    \item We propose \ModelName, a re-ranking framework that applies GNN-based propagation over query-induced subgraphs to capture cross-document interactions at $\mathcal{O}(c \cdot K)$ online cost (linear in $K$ for fixed graph degree $c$), comparing favorably to the $\mathcal{O}(K^2)$ complexity of self-attention-based re-rankers.
    \item We introduce a mechanism to extract query-specific subgraphs from a large pre-computed corpus graph, enabling efficient offline-online separation.
    \item We provide a comprehensive empirical comparison of five GNN architectures against self-attention baselines across three TREC benchmarks with varying difficulty profiles. GNN-based re-ranking is uniquely robust on the hardest benchmark (DLHard), outperforming self-attention re-ranking by up to $+9.0\%$ AP while requiring linear rather than quadratic computation. We additionally report a detailed inference-time efficiency analysis.
\end{itemize}

\begin{figure}[p]
    \centering
    \includegraphics[width=\textwidth]{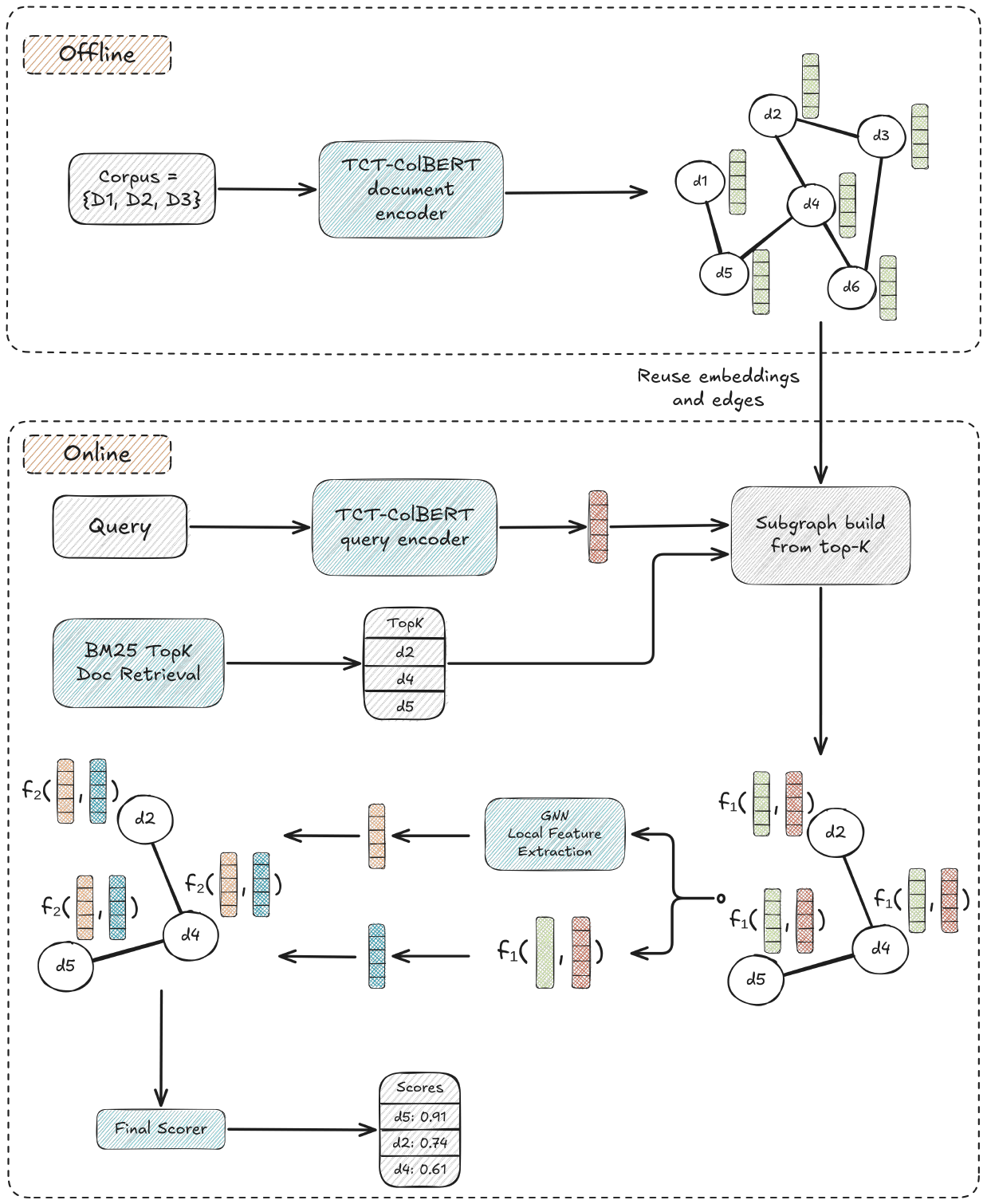}
    \caption{\ModelName pipeline ($K=3$ example). \textit{Offline}, TCT-ColBERT encodes the entire corpus and connects each document to its $c$ nearest semantic neighbors, building corpus graph $\mathcal{G}$ once for reuse across all queries. At query time, BM25 retrieves the top-$K$ candidates and TCT-ColBERT encodes the query; the candidate set determines a query-induced subgraph of $\mathcal{G}$, reusing offline document embeddings, while the query embedding flows into node feature construction. Concretely, $f_1$ combines the query and document embeddings into a per-candidate node feature; a GNN then propagates cross-document signals over the subgraph; and $f_2$ merges the GNN output with the original node feature before a final linear scorer produces the ranking.}
    \label{fig:overview}
\end{figure}

\section{Related Work}
Transformer-based re-rankers such as MonoBERT~\cite{monobert}, ColBERT~\cite{colbert}, and TCT-ColBERT~\cite{tctcolbert} produce high-accuracy relevance scores but treat candidates independently.

\paragraph{Set-based re-ranking.}
SetRank~\cite{pang2020setrank} and related works~\cite{pobrotyn2020context,pasumarthi2020permutation} apply self-attention over the full candidate set to capture inter-document dependencies. While effective, these methods incur $\mathcal{O}(K^2)$ cost and, as we show, can be fragile on harder query sets where relevant documents are sparsely distributed.

\paragraph{Graph-based representations in IR.}
PageRank~\cite{page1999pagerank} exemplifies how document link structure can propagate importance signals. More recently, MacAvaney et al.~\cite{corpusgraph} proposed Graph-based Adaptive Re-ranking (GAR), a retrieval framework that leverages a pre-computed semantic corpus graph to expand the initial candidate pool. GAR traverses the graph to add documents that are semantically related to the initially retrieved candidates, substantially improving retrieval recall.

\ModelName differs fundamentally: GAR augments the \emph{retrieval} stage, while \ModelName uses corpus graph structure during \emph{re-ranking}.

GNNs address this gap by jointly learning representations and propagating relevance signals across graph structure~\cite{messagepassing,everythingisconnected}. GNNs have been applied to ranking in several settings: custom PageRank from labeled examples~\cite{scarselli2005graph}, global ranking from pairwise comparisons~\cite{damke2021ranking,he2022gnnrank}, node importance in networks~\cite{liu2022learning,qu2023gnr}, and web resource ranking~\cite{ergashev2023learning}. Intra-document word graphs have also been used for relevance matching~\cite{graphofwords,graphofwords2}. Cross-document structure for re-ranking has been explored in multi-hop QA~\cite{nie2021answeringanyhopopendomainquestions} where graphs are dynamically constructed per-question over reasoning chains; differently, we target general-purpose passage re-ranking using a static, pre-computed corpus graph built once offline and reused across queries. See~\cite{zaoad2025graphbasedrerankingemergingtechniques} for a broader survey of graph-based re-ranking approaches. To our knowledge, no prior work applies GNN-based propagation over query-induced subgraphs from a pre-computed corpus graph for general-purpose document re-ranking.

\section{Methodology}
\label{sec:methodology}

\ModelName incorporates document-document relationships into neural re-ranking through three stages: (i) offline corpus graph construction, (ii) retrieval and query-induced subgraph extraction, and (iii) GNN-based re-ranking. Figure~\ref{fig:overview} illustrates the full pipeline.

\subsubsection*{Corpus Graph}
Given a document corpus $\mathcal{C}$, we construct an offline semantic corpus graph $\mathcal{G} = (\mathcal{C}, \mathcal{E})$ where each document $d \in \mathcal{C}$ is a node and edges $(d_i,d_j) \in \mathcal{E}$ connect each document to its $c$ nearest neighbors by cosine similarity in the TCT-ColBERT embedding space, ensuring fixed-degree sparsity. Although this offline construction is computationally expensive, it is performed only once and amortized across all queries.
We follow the construction procedure of~\cite{corpusgraph}.

\subsubsection*{Retrieval and Subgraph Extraction}
Given a query $q$, BM25~\cite{bm25} retrieves the top-$K$ documents $\mathcal{C}_q \subseteq \mathcal{C}$. From $\mathcal{G}$, we extract the query-induced subgraph retaining only retrieved nodes and edges between them:
\begin{equation} \label{eq:subgraph}
    \mathcal{G}_{q} = (\mathcal{C}_q, \mathcal{E}_q), \quad \mathcal{E}_q = \{(d_i,d_j) \in \mathcal{E} \;|\; d_i,d_j \in \mathcal{C}_q\}.
\end{equation}
Connectivity is bounded by construction: $|\mathcal{E}_q| \leq c \cdot K$. Each node $d_i$ is assigned feature vector $\mathbf{x}_i = f_1(\mathbf{z}_{q}, \mathbf{z}_{d_i})$, with $f_1 = \odot$ (element-wise product) and $\mathbf{z}_{q}, \mathbf{z}_{d_i} \in \mathbb{R}^m$ denoting query and document embeddings from TCT-ColBERT. The resulting feature matrix $\mathbf{X}_q \in \mathbb{R}^{|\mathcal{C}_q| \times m}$ encodes query-document relevance per node.

\subsubsection*{Re-Ranking}
The GNN operates over $\mathcal{G}_q$ to propagate cross-document signals. Each feature vector is augmented with the document's BM25 score to retain lexical relevance as an explicit feature. The input $\mathbf{X}'_q \in \mathbb{R}^{|\mathcal{C}_q| \times (m+1)}$ is processed by the GNN:
\begin{equation}
\label{eq:local_interactions}
    \mathbf{H}_{q,\text{GNN}} = \text{GNN}(\mathbf{X'}_q, \mathbf{A}_q),
\end{equation}
where $\mathbf{A}_q$ is the adjacency matrix of $\mathcal{G}_q$ and the hidden dimension is $m'$. The GNN output is merged with the original features by concatenation ($f_2 = \|$):
\begin{equation}
    \mathbf{H}_q = f_2\left(\mathbf{H}_{q,\text{GNN}}, \mathbf{X}_q\right).
\end{equation}
A linear scorer $\phi_{\theta}$ maps each row of $\mathbf{H}_q$ to a scalar relevance score, producing the final re-ranking:
\begin{equation}
    \pi_q = \text{ARGSORT}(\phi_{\theta}(\mathbf{H}_q)).
\end{equation}

\paragraph{Complexity.} By construction $|\mathcal{E}_q| \leq c \cdot K$, so GNN message passing costs $\mathcal{O}(c \cdot K)$ for fixed corpus graph degree $c$. This scales linearly with $K$, compared to $\mathcal{O}(K^2)$ for self-attention over the same candidate set.

\section{Experimental Setup}
We address three research questions:
\begin{itemize}[leftmargin=1.1cm]
\item[\textbf{RQ1.}] Does modeling cross-document interactions via GNNs improve re-ranking over a strong transformer baseline?
\item[\textbf{RQ2.}] How does GNN-based re-ranking compare to self-attention re-ranking in terms of effectiveness and efficiency?
\item[\textbf{RQ3.}] How much does the GNN branch itself contribute to the observed gains?
\end{itemize}

\subsection{Datasets}
We use the MS MARCO passage ranking corpus~\cite{msmarco} (8.8M documents). We use 14,994 queries for training and 200 queries for validation, both drawn from the MS MARCO training split. We evaluate on three standard test sets: TREC-DL19, TREC-DL20~\cite{dl19-20}, and TREC-DLHard~\cite{dl-hard}.
\begin{itemize}
    \item \textbf{DL19}: 43 queries, average 215 relevance judgments per query.
    \item \textbf{DL20}: 54 queries, average 211 relevance judgments per query.
    \item \textbf{DLHard}: 50 difficult queries, partially relabeled from DL19/DL20, targeting topics that challenge neural ranking models.
\end{itemize}

\subsection{Models}
We compare the following pipelines:
\begin{itemize}
    \item \textbf{BM25}~\cite{bm25}: lexical-based retrieval baseline.
    \item \textbf{TCT-ColBERT}~\cite{tctcolbert}: transformer-based re-ranker on top of BM25 (primary baseline).
    \item \textbf{Self-Attention}: a multi-layer Transformer encoder applied to the full BM25 candidate set features (same $f_1$ as \ModelName), trained with LambdaRank. This is an $\mathcal{O}(K^2)$ set-based re-ranker and serves as the direct efficiency comparison to \ModelName.
    \item \textbf{GAR}~\cite{corpusgraph}: graph-augmented retrieval using the same pre-computed corpus graph to expand BM25 candidates. GAR operates at the \emph{retrieval} stage (not re-ranking) and is included as a reference for corpus-graph-aware performance.
    \item \textbf{GNRR variants}: \ModelName instantiated with GCN~\cite{GCN}, GraphSAGE~\cite{GraphSAGE}, GAT~\cite{GAT}, GIN~\cite{GIN}, and SignedConv~\cite{signedconv} as the GNN operator, all trained over the semantic corpus graph.
\end{itemize}
We report four standard IR metrics: Average Precision (AP), Reciprocal Rank (RR), Precision at rank 3 (P@3), and nDCG@10 (nG@10) with relevance$\geq 2$.

\subsection{Training}
All variants are trained with LambdaRank~\cite{lambdarank}. Training runs for up to 100 epochs with early stopping on validation nDCG@10 (patience 7). The TCT-ColBERT encoder remains frozen. We report results averaged over 3 random seeds.

\subsection{Hyperparameters}
We set the corpus graph neighborhood size to $c = 16$ and $K = 1000$. TCT-ColBERT produces embeddings of dimension $m = 768$. We tune $L \in \{1, 2, 3\}$ and $m' \in \{64, 128\}$ on the validation set. All the hyperparameters and model checkpoints are available in our code repository.


\begin{table}[ht]
    \centering
    \footnotesize
    \caption{Re-ranking on TREC-DL19, DL20, and DLHard using TCT-ColBERT features ($K=1000$). Self-Attention and all \ModelName variants are independent re-rankers applied on top of frozen TCT-ColBERT representations. Best score per column in \textbf{bold}. $^*$ and $^{**}$ indicate a significant t-test at $0.1$ and $0.05$ levels respectively.}
    \vspace{0.1cm}
    \setlength{\tabcolsep}{3pt}
    \resizebox{\textwidth}{!}{%
    \begin{tabular}{lcccccccccccc}
        \toprule
        \multirow{2}{*}{\textbf{Pipeline}}
            & \multicolumn{4}{c}{\textbf{DL19}}
            & \multicolumn{4}{c}{\textbf{DL20}}
            & \multicolumn{4}{c}{\textbf{DLHard}} \\
        \cmidrule(lr){2-5} \cmidrule(lr){6-9} \cmidrule(lr){10-13}
            & AP & RR & P@3 & nG@10
            & AP & RR & P@3 & nG@10
            & AP & RR & P@3 & nG@10 \\
        \midrule
        BM25
            & 0.286 & 0.642 & 0.473 & 0.480
            & 0.293 & 0.619 & 0.463 & 0.494
            & 0.147 & 0.422 & 0.240 & 0.274 \\
        \addlinespace[2pt]
        +TCT-ColBERT
            & 0.430 & 0.843 & 0.667 & 0.685
            & 0.453 & 0.817 & 0.691 & 0.680
            & 0.230 & 0.538 & 0.353 & 0.373 \\
        \cmidrule(l){1-13}
        \addlinespace[1pt]
        \quad+Self-Attention
            & 0.431 & 0.844 & 0.674 & 0.686
            & 0.454 & 0.827 & \textbf{0.704} & 0.684
            & 0.222 & 0.529 & 0.340 & 0.368 \\
        \quad+GCN
            & \textbf{0.455}$^{**}$ & \textbf{0.858} & 0.713 & \textbf{0.702}$^*$
            & \textbf{0.470}$^{**}$ & \textbf{0.840} & 0.673 & \textbf{0.695}$^*$
            & \textbf{0.242} & \textbf{0.559} & \textbf{0.367} & \textbf{0.386} \\
        \quad+GraphSAGE
            & 0.434 & 0.850 & \textbf{0.729} & 0.689
            & 0.454 & 0.837 & 0.667 & 0.685
            & 0.223 & 0.531 & \textbf{0.367} & 0.379 \\
        \quad+GAT
            & 0.441 & 0.852 & 0.698 & 0.690
            & 0.453 & 0.826 & 0.685 & 0.682
            & 0.219 & 0.542 & \textbf{0.367} & 0.376 \\
        \quad+GIN
            & 0.449$^*$ & 0.853 & \textbf{0.729} & 0.691
            & 0.455 & 0.793 & 0.642 & 0.675
            & 0.215 & 0.490 & 0.333 & 0.357 \\
        \quad+SignedConv
            & 0.426 & 0.828 & 0.698 & 0.675
            & 0.445 & 0.813 & 0.685 & 0.676
            & 0.218 & 0.509 & 0.353 & 0.366 \\
        \bottomrule
    \end{tabular}%
    }
    \label{tab:benchmarking}
\end{table}

\begin{table}[ht]
    \centering
    \footnotesize
    \caption{Effect of improved retrieval recall on re-ranking. GAR augments BM25 candidates with semantic-graph-neighbor documents; \ModelName re-rankers (trained on BM25 pools) are applied to the GAR-enlarged pool without retraining, using the same semantic corpus graph as during training. Best per column in \textbf{bold}. 
    }
    \vspace{0.1cm}
    \setlength{\tabcolsep}{3pt}
    \resizebox{\textwidth}{!}{%
    \begin{tabular}{lcccccccccccc}
        \toprule
        \multirow{2}{*}{\textbf{Pipeline}}
            & \multicolumn{4}{c}{\textbf{DL19}}
            & \multicolumn{4}{c}{\textbf{DL20}}
            & \multicolumn{4}{c}{\textbf{DLHard}} \\
        \cmidrule(lr){2-5} \cmidrule(lr){6-9} \cmidrule(lr){10-13}
            & AP & RR & P@3 & nG@10
            & AP & RR & P@3 & nG@10
            & AP & RR & P@3 & nG@10 \\
        \midrule
        BM25
            & 0.286 & 0.642 & 0.473 & 0.480
            & 0.293 & 0.619 & 0.463 & 0.494
            & 0.147 & 0.422 & 0.240 & 0.274 \\
        +TCT-ColBERT
            & 0.430 & 0.843 & 0.667 & 0.685
            & 0.453 & \textbf{0.817} & \textbf{0.691} & 0.680
            & 0.230 & \textbf{0.538} & 0.353 & 0.373 \\
        \cmidrule(l){1-13}
        \addlinespace[1pt]
        GAR
            & 0.440 & 0.833 & 0.667 & 0.696
            & 0.447 & 0.815 & 0.685 & 0.680
            & \textbf{0.233} & 0.532 & 0.353 & \textbf{0.373} \\
        \addlinespace[2pt]
        \quad+GCN
            & 0.427 & 0.816 & 0.674 & 0.698
            & 0.443 & 0.804 & 0.642 & 0.666
            & 0.219 & 0.503 & 0.340 & 0.357 \\
        \quad+GraphSAGE
            & 0.431 & 0.815 & 0.698 & 0.697
            & 0.432 & 0.790 & 0.667 & 0.664
            & 0.220 & 0.501 & \textbf{0.360} & 0.356 \\
        \quad+GAT
            & \textbf{0.444} & 0.833 & \textbf{0.721} & \textbf{0.705}
            & \textbf{0.458} & 0.808 & 0.679 & \textbf{0.681}
            & 0.229 & 0.495 & \textbf{0.360} & 0.368 \\
        \quad+GIN
            & 0.429 & 0.790 & 0.674 & 0.676
            & 0.421 & 0.777 & 0.617 & 0.616
            & 0.200 & 0.456 & 0.320 & 0.325 \\
        \quad+SignedConv
            & 0.419 & \textbf{0.844} & 0.682 & 0.686
            & 0.413 & 0.788 & 0.642 & 0.654
            & 0.212 & 0.498 & 0.353 & 0.348 \\
        \bottomrule
    \end{tabular}%
    }
    \label{tab:gar_gnrr}
\end{table}

\section{Results}

\subsection{GNN Re-Ranking vs.\ TCT-ColBERT (RQ1)}

Table~\ref{tab:benchmarking} shows that GNN re-ranking improves over TCT-ColBERT for the most effective architectures, with variation across models and benchmarks. GCN is the only architecture that improves AP across all three benchmarks: $+5.8\%$ on DL19, $+3.8\%$ on DL20, $+5.2\%$ on DLHard, with consistent nDCG@10 gains of $+2.5\%$, $+2.2\%$, and $+3.5\%$ respectively. GraphSAGE improves AP on DL19 ($+0.9\%$) but degrades AP on DLHard ($-3.0\%$); it improves nDCG@10 ($+1.6\%$) and P@3 ($+4.0\%$) on DLHard. GAT gains $+2.6\%$ AP on DL19 but degrades AP on DLHard ($-4.8\%$), while improving nDCG@10 ($+0.8\%$) and P@3 ($+4.0\%$) there. GIN gains $+4.4\%$ AP on DL19 but degrades on DLHard ($-6.5\%$ AP). SignedConv falls below TCT-ColBERT across all three benchmarks.

On DLHard, GCN is the only architecture that improves AP, raising it from $0.230$ to $0.242$ ($+5.2\%$) alongside nDCG@10 gains of $+3.5\%$. GAT improves nDCG@10 ($+0.8\%$) and P@3 ($+4.0\%$) on DLHard but not AP ($-4.8\%$); GIN degrades across all DLHard metrics ($-6.5\%$ AP); SignedConv degrades AP, RR, and nDCG@10 on DLHard, with P@3 flat at $0.353$. These results suggest that the choice of GNN aggregation scheme substantially influences generalization to harder query distributions.

\subsection{GNN vs.\ Self-Attention (RQ2)}

\paragraph{Effectiveness.}
Table~\ref{tab:benchmarking} reveals an asymmetry between GCN and self-attention re-ranking across benchmarks.
On DL19, self-attention improves by $+0.2\%$ AP over TCT-ColBERT; GCN achieves $+5.8\%$ AP on the same benchmark.
On DL20, self-attention gains $+0.2\%$ AP; GCN gains $+3.8\%$ AP, with smaller gains from other variants.
On DLHard, self-attention degrades TCT-ColBERT by $-3.5\%$ AP ($0.222$ vs.\ $0.230$) and $-1.3\%$ nDCG@10. GCN is the only method that improves AP on DLHard, achieving $+5.2\%$ over TCT-ColBERT and $+9.0\%$ over self-attention ($0.242$ vs.\ $0.222$).

We attribute this asymmetry to structural regularization: the sparse corpus graph constrains information flow to semantically similar document pairs, while the full $K\!\times\!K$ attention matrix over predominantly non-relevant candidates amplifies noise in hard-query settings. The GNN restricts information flow to corpus-graph neighbors, filtering irrelevant cross-document interactions by construction.

\paragraph{Inference efficiency (Table~\ref{tab:efficiency}).}
\begin{table}[ht]
    \centering
    \footnotesize
    \caption{Inference efficiency at $K=1000$. Latency averaged over DL19 queries on NVIDIA RTX 3090 Ti (GNN forward pass + subgraph extraction; TCT encoding excluded). GNN methods cost $\mathcal{O}(c \cdot K)$ with $c=16$, versus $\mathcal{O}(K^2)$ for self-attention.}
    \vspace{0.1cm}
    \setlength{\tabcolsep}{4pt} 
    \begin{tabular}{lrrr}
        \toprule
        \textbf{Method} & \textbf{Complexity} & \textbf{Params} & \textbf{ms/query} \\
        \midrule
        Self-Attention       & $\mathcal{O}(K^2)$         & 363.9K & 37.3 \\
        \midrule
        GNRR+GCN       & $\mathcal{O}(c \cdot K)$   & 279.6K & 34.2 \\
        GNRR+GraphSAGE & $\mathcal{O}(c \cdot K)$   & 160.2K & 32.8 \\
        GNRR+GAT       & $\mathcal{O}(c \cdot K)$   & 246.8K & 34.3 \\
        GNRR+GIN       & $\mathcal{O}(c \cdot K)$   & 123.5K & 32.4 \\
        GNRR+SignedConv & $\mathcal{O}(c \cdot K)$  & 295.7K & 35.2 \\
        \bottomrule
    \end{tabular}
    \label{tab:efficiency}
\end{table}

GNN models are faster and smaller than self-attention: at $K=1000$, GNN latency is $5$--$13\%$ lower (32--35~ms vs.\ 37.3~ms) with up to $66\%$ fewer parameters (GIN: 124K vs.\ 364K). The advantage grows with $K$: self-attention cost scales as $\mathcal{O}(K^2)$, so doubling $K$ quadruples its cost while GNN cost grows as $\mathcal{O}(c \cdot K)$ and remains linear. GCN achieves better DLHard effectiveness ($+5.2\%$ AP) at lower computational cost than self-attention.

\subsection{Sensitivity to Candidate Distribution}\label{sec:gar}
All experiments presented so far use candidate pools initially retrieved by BM25. To assess whether GNRR depends on the distribution of BM25 candidates and to evaluate its complementarity with retrieval-enhancement methods such as GAR, we conduct an additional experiment.  \cite{corpusgraph} introduced the semantic corpus graph as a mechanism for increasing the recall of the top-1000 retrieved candidates. The key intuition is that documents that are semantically similar to relevant documents may themselves be relevant. Their results show that improving candidate recall at the retrieval stage can enhance downstream re-ranking performance by recovering relevant documents that BM25 fails to retrieve.

Table~\ref{tab:gar_gnrr} examines what happens when \ModelName re-rankers (trained on BM25 candidate pools) are applied to GAR-augmented pools without retraining, using the same semantic corpus graph as during training.
GAR raises DL19 AP from $0.430$ to $0.440$ ($+2.3\%$) and nDCG@10 from $0.685$ to $0.696$ over the BM25+TCT-ColBERT baseline through retrieval-stage candidate expansion alone.

The five \ModelName architectures respond differently to this distribution shift.
GCN, GIN, GraphSAGE, and SignedConv all degrade AP relative to GAR alone across all three benchmarks, consistent with out-of-distribution failure: these models were trained to re-rank BM25 candidate pools and their fixed, non-adaptive aggregation schemes do not transfer directly to the larger, more diverse candidate distribution produced by GAR.
GAT is the exception on DL19 and DL20. Its attention mechanism can adaptively reweight neighbors at test time, improving AP over GAR alone by $+0.9\%$ on DL19 ($0.444$ vs.\ $0.440$) and $+2.5\%$ on DL20 ($0.458$ vs.\ $0.447$); GAT also records the best nDCG@10 on both benchmarks. On DLHard, all five architectures degrade, suggesting that the shift in candidate distribution is particularly challenging on harder queries regardless of aggregation scheme.

These results carry two messages. First, attention-based aggregation generalizes more broadly across candidate distributions than fixed isotropic GNN operators. Second, retrieval and re-ranking stages must be jointly trained when the candidate source changes; re-training \ModelName on GAR-augmented pools is a natural direction for future work.
\begin{figure}[t]
  \centering
  \includegraphics[width=\textwidth]{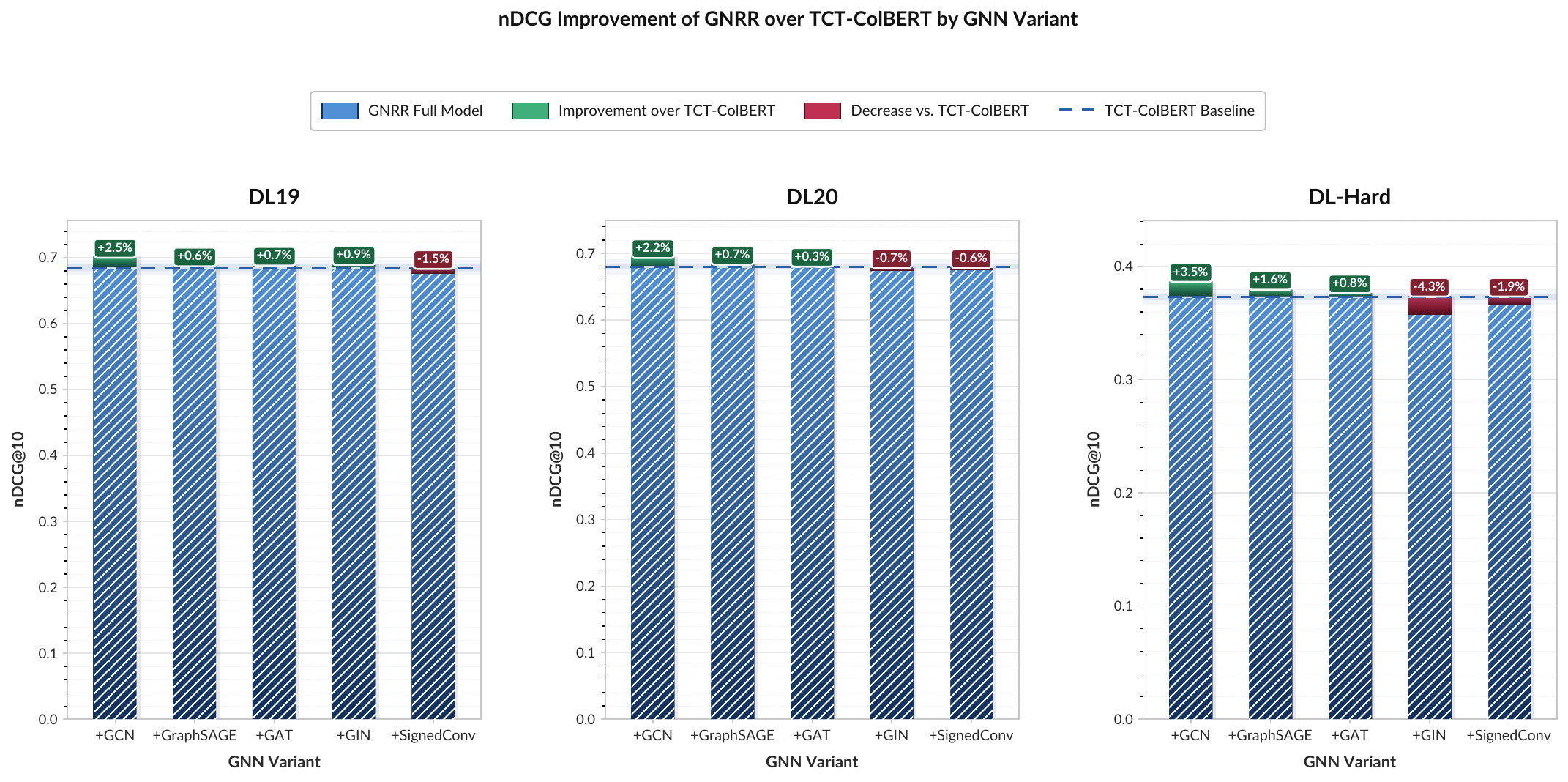}
    \caption{nDCG@10 improvement of each \ModelName variant over the TCT-ColBERT baseline on DL19, DL20, and DLHard. Green caps denote gains; bordeaux caps denote degradation. The dashed line marks the baseline (zero improvement).}
  \label{fig:ndcg_ablation}
\end{figure}
\subsection{GNN Branch Contribution (RQ3)}
To isolate the contribution of graph propagation, we perform an ablation in which the GNN branch output $\mathbf{H}_{q,\text{GNN}}$ is replaced by a zero vector at inference time, removing all graph-derived information while leaving the rest of the model unchanged.

Figure~\ref{fig:ndcg_ablation} reports the relative change in nDCG@10 when the GNN branch is reactivated. GCN yields the largest and most consistent gains, improving nDCG@10 by +2.5\% on DL19, +2.2\% on DL20, and +3.5\% on DLHard. GraphSAGE and GAT provide smaller improvements, while SignedConv offers little benefit.

These results indicate that graph-based message passing contributes useful ranking information beyond individual query-document representations. The particularly strong gains on DLHard suggest that cross-document signals are most valuable for difficult queries.

\section{Conclusions}
This paper introduced \ModelName, a re-ranking framework that captures cross-document interactions by extracting a sparse, query-induced subgraph from a pre-computed semantic corpus graph and applying a GNN over transformer-based representations. Unlike self-attention re-rankers with $\mathcal{O}(K^2)$ cost, \ModelName requires only $\mathcal{O}(c \cdot K)$ online computation, linear in the candidate set size for fixed corpus graph degree $c$.

Comparison with self-attention re-ranking reveals that GCN is the only method in our evaluation that improves AP on harder query sets: on TREC-DLHard, GCN achieves $+5.2\%$ AP over TCT-ColBERT and $+9.0\%$ AP over self-attention, whereas self-attention re-ranking \emph{degrades} below the TCT-ColBERT baseline on the same set. Efficiency analysis shows that GCN achieves these gains with fewer parameters and lower per-query latency at $K=1000$ than self-attention, with a theoretical linear vs.\ quadratic scaling advantage as $K$ grows.

These results suggest that sparse inter-document structure via pre-computed corpus graphs is a viable and computationally efficient mechanism for improving neural re-ranking, with particular value for difficult retrieval scenarios.

\subsection*{Limitations and Future Work}
Results across architectures are mixed on DL20: GCN gains $+3.8\%$ AP while most other variants are flat or below TCT-ColBERT, suggesting that the benefit of graph propagation depends on the aggregation scheme. Future work includes: (1) heterogeneous graph structures encoding multiple relation types; (2) joint training with the dense encoder; (3) adaptive graph construction per query; and (4) scaling the corpus graph neighborhood size and studying its interaction with re-ranking quality. The gap between GAR and \ModelName (re-ranking) also suggests that combining the two approaches through joint training is a natural direction.

\begin{credits}
\subsubsection{Generative AI Usage Disclosure}
Generative AI tools were used for debugging assistance and language editing only. Conceptualization, methodology, experimental design, analysis, and manuscript writing were conducted entirely by the authors.

\subsubsection{Funding}

This work was supported by the EVOLVE project, ``A Fluid Framework for Dynamic Agentic AI Systems'' (Sapienza Università di Roma), and partially supported by the FoReLab and CrossLab projects (Departments of Excellence) and by the NVIDIA Academic Grants Program 2026.
This research was conducted as part of the ENRESFOOD project, funded under the FutureFoodS partnership. The project is co-funded by the Austrian Science Fund (FWF; 10.55776/KIN4661525), The Dutch Ministry of Agriculture, Fisheries, Food Security and Nature (BO-43-219-026), the Italian Ministry of University and Research (MUR), the German Federal Ministry of Research, Technology and Space (031B1717), and the European Union’s Horizon Europe research and innovation programme via FutureFoodS, the European Partnership for a Sustainable Future of Food Systems, co-funded under Grant Agreement No. 101136361.

\subsubsection{\discintname}
The authors have no competing interests to declare that are relevant to the content of this article.
\end{credits}

%
%
\bibliographystyle{splncs04}
\bibliography{main}

\end{document}